\documentclass[10pt]{revtex4}
\usepackage{hyperref}
\usepackage{color} 
\begin{document}

\title{$N=1$ super-Chern-Simons theory in Batalin-Vilkovisky formulation}

\author{ Sudhaker Upadhyay}
 \email {  sudhakerupadhyay@gmail.com; 
 sudhaker@boson.bose.res.in}

\affiliation { S. N. Bose National Centre for Basic Sciences,\\
Block JD, Sector III, Salt Lake, Kolkata -700098, India. }

\begin{abstract}
  We analyse the  Abelian $N=1$ super-Chern-Simons model
coupled to parity-preserving matter in linear and non-linear gauges with
exact BRST invariance. Then we analyse the theory in field/antifield formulation to discuss the model at quantum level. Furthermore, we implement the field/antifield dependent transformation parameter to generalize the BRST symmetry of the theory. The novelty of field/antifield dependent BRST transformation is that under change of variable the Jacobian of the 
functional integral extends the quantum action from linear gauge to non-linear gauge. The results are established in full generality.
  \end{abstract}
\maketitle

\section{  Introduction}
The supersymmetric  as well as the ordinary version  of Abelian gauge field theories in three 
dimensional (3D) space-times
are   subject of enormous interest in a recent
past \cite{al,sd,ma}.  At low energies, supersymmetric Chern-Simons theories are important because they  describe the world-volume of $M2$-membranes in
$M$-theory  \cite{gus,bag,bag1,bag2,bag3}.  In fact, the world-volume of
$M2$-membranes in $M$-theory, at low energies, is thought to be described by the $N= 8$ superconformal Chern-Simons-matter theory \cite{bis}.  However, Chern-Simons theory with
$N= 1$ supersymmetry has also been studied
in relation to axion gauge symmetry \cite{bis1}. Besides
their relevance in connection with the possibility of getting non-perturbative results, the
recent results on the Landau gauge finiteness of Chern-Simons theories are  remarkable  
that make  3D gauge theories
so attractive    \cite{fd,ab,cl}. The 
gauge theories in 3D space-times  provide a basis to tackle exciting topics of condensed matter physics such as high temperature
superconductivity and fractional quantum hall effect \cite{js}.
It is well known that the Landau gauge has very special features  as compared to generic
linear or non-linear gauge-fixings \cite{naka,lu}.
 These gauge  conditions can be incorporated to the theory  
at quantum level by adding the suitable  gauge-fixing and ghost terms to the classical action which remains  invariant under the fermionic rigid
 BRST  invariance \cite{brst}.
 However, the BRST symmetry plays an  important role in proof of  unitarity and renormalizability of the gauge theories \cite{ht,wei}.  

On the other hand, the Batalin-Vilkovisky (BV) formulation is a more fundamental approach  to quantize the more general 
gauge theories with open gauge algebra such as the supergravity   and topological field theories \cite{ht,wei,bv,bv1,bv2,subm, bss}.
Subsequently, the  generalization of BRST transformations by making the infinitesimal parameter finite and field dependent is originally advocated in \cite{sdj} and 
 has found several applications in gauge field theories 
\cite{sdj1,rb,susk,sb,smm,fs,sud1,sudhak, sudhak1, rs}.

Our aim is to connect the different solutions (extended quantum actions) of the quantum master equation for the $N=1$ super-Chern-Simons theory.
We establish the results which hold up to all order of corrections.
 For this purpose, we first investigate the theory in linear and non-linear gauges with their
 BRST invariances. Furthermore, the BRST symmetry transformation is made field/antifield dependent.
 The Jacobian of the path integral measure is evaluated explicitly under change of variables.
 We show that the solutions of the quantum master equation in BV  formulation 
can be mapped under field/antifield dependent BRST transformation. 
 
The present paper is organized as follows. In the second section, we discuss the
$N=1$ super-Chern-Simons model in linear and non-linear gauges possessing BRST symmetry.
In the third section, we analyse the theory in the
 field/antifield formulation which describes the theory
at  quantum level. Solutions of quantum master equations are connected through field/antifield
dependent BRST transformation in the fourth section. The last section summarizes our
results.

\section{ The model and the BRST symmetries}
In this section, we analyse the $N=1$ Abelian super-Chern-Simons theory with their supersymmetric
BRST invariance. For this purpose let us start with the gauge-invariant action for the $N=1$ super-Chern-Simons theory coupled  with matter supermultiplets in a parity-preserving
way,  in superspace,  given by \cite{cim,cim1,cim2}
\begin{eqnarray}
\Sigma _{inv}  &=&  \frac{1}{2} \int dv\left[ k( \Gamma^\alpha W_\alpha)-(
\nabla^\alpha \bar{\Phi} _{+})(\nabla_\alpha \Phi_{+})-(\nabla^\alpha \bar{\Phi}
_{-})(\nabla_\alpha \Phi_{-})+m(\bar{\Phi} _{+}\Phi_{+}-\bar{\Phi} _{-}\Phi_{-})  
\right.\nonumber \\  
&-&\left.   (\bar{\Phi}_{+}\Phi_{+}-\bar{\Phi}_{-}\Phi_{-})^2 \right],
\label{inv} 
\end{eqnarray}
where the spinorial Majorana superfield $\Gamma_\alpha$ is the gauge superconnection and matter is represented by the complex scalar superfields $\Phi_\pm$  with opposite $U(1)$-charges.
  Here the real parameter $k$  has dimension of mass and superspace measure $dv$ has following expression: $dv=d^3xd^2\theta$.
The covariant spinorial derivatives  for $\Phi_\pm$ and their conjugate superfields $\bar\Phi_\pm$  are defined as  
\begin{eqnarray}
\nabla_\alpha\Phi_{\pm}=\left( D_\alpha \mp i\Gamma_\alpha\right) \Phi_{\pm} \ \
\mbox{and}\ \  \nabla_\alpha \bar{\Phi}_{\pm}=\left({D}_\alpha\pm i{
\Gamma}_\alpha\right) \bar{\Phi}_{\pm},  \label{covder}
\end{eqnarray}
where spinorial derivative has the following expression: $D_\alpha =\partial_\alpha+i\theta^\beta\partial_{\alpha\beta}$. 

Furthermore, we define the superfield-strength for gauge superconnection as
\begin{equation}
W_\alpha=\frac{1}{2}D^\beta D_\alpha \Gamma_\beta.
\end{equation}
 The components of superfields $\Gamma, \Phi_\pm$ and $\bar\Phi_\pm$ are defined  in terms of
  spinor derivatives $D_\alpha$
as follows:
\begin{eqnarray}
&&\chi_\alpha =\Gamma_\alpha|, \ \ \ \ \ \ B=\frac{1}{2}D^\alpha\Gamma_\alpha|,\nonumber\\
&&V_{\alpha\beta}=-\frac{i}{2}D_{(\alpha}\Gamma_{\beta)}|,\ \ \lambda_\alpha =\frac{1}{2}D^\beta D_\alpha \Gamma_\beta|,
\nonumber\\
&&A_{\pm}(x)=\Phi_{\pm}(x, \theta)|,\ \ \ \ \ \bar A_{\pm}(x)=\bar\Phi_{\pm}(x, \theta)|,\nonumber\\
&&\psi^\alpha_{\pm}(x)=D^\alpha\Phi_{\pm}(x, \theta)|,\ \ \bar \psi^\alpha_{\pm}(x)=D^\alpha\bar\Phi_{\pm}(x, \theta)|,\nonumber\\
&&F_{\pm}(x)=D^2\Phi_{\pm}(x, \theta)|,\ \ \ \ \bar F_{\pm}(x)=D^2\bar\Phi_{\pm}(x, \theta)|,
\end{eqnarray}
where ``$|$" denotes the quantity   evaluated at $\theta=0$. The gauge invariance 
of the $N=1$ Abelian super-Chern-Simons  action given in (\ref{inv}) \cite{cim,cim1,cim2} reflects the redundancy
in gauge degree of freedoms. 
Therefore, one needs to break the 
local gauge invariance to quantize the theory correctly. 
There may be many choices for the gauge condition as the physical theory 
does not depend on the choices of the gauge condition \cite{ht}. 
For the present analysis, we  choose the following  well established linear (Landau) gauge condition:
\begin{equation}
\Omega_1 :=D^\alpha\Gamma_\alpha  =0.
\end{equation}
This gauge condition can be employed to the theory at quantum level by adding the following 
 gauge-fixing and ghost terms in the action:
\begin{eqnarray}
\Sigma^L_{gf+gh} =  \int dv\left[ BD^\alpha\Gamma_\alpha +{ \widehat C}D^2 C\right].  
\label{gf} 
\end{eqnarray}
Now, the total effective action in Landau gauge reads
\begin{equation}
\Sigma_{eff}^L = \Sigma_{inv}+\Sigma^L_{gf+gh},\label{effl}
\end{equation}
which   remains  invariant under  the
following nilpotent BRST transformations:
\begin{eqnarray}
&&\delta_b\Phi_{\pm}=\pm iC\Phi_{\pm}\eta,\ \  \ \delta_b\bar\Phi_{\pm}=\mp iC\bar\Phi_{\pm}\eta,\nonumber\\
&&\delta_b\Gamma_\alpha =D_\alpha C\eta,\ \ \ \ \ \ \delta_b\widehat{C}=B\eta,\nonumber\\
&&\delta_bC=0,\ \ \ \ \ \ \  \delta_b B=0,\label{brs}
\end{eqnarray}
where the transformation parameter $\eta$ is Grassmannian in nature.
Furthermore, we restrict the gauge superfield to satisfy another 
 gauge condition which is non-linear (quadratic) in nature as follows
\begin{equation}
\Omega_2 :=D^\alpha\Gamma_\alpha +\beta\Gamma_\alpha\Gamma^\alpha =0,
\end{equation}
where $\beta$ is an arbitrary constant.
For this gauge choice the gauge-fixing and ghost terms  can be written as
\begin{eqnarray}
\Sigma^{NL} _{gf+gh} =  \int dv\left[ B(D^\alpha\Gamma_\alpha +\beta\Gamma_\alpha\Gamma^\alpha) +{ \widehat C}D^2 C +2\beta{\widehat C}\Gamma_\alpha D^\alpha  C \right].  
\label{ngf} 
\end{eqnarray}
The effective action for the $N=1$ Abelian super-Chern-Simons theory in such non-linear gauge 
is given by
 \begin{equation}
\Sigma_{eff}^{NL}= \Sigma_{inv}+\Sigma^{NL}_{gf+gh},
\end{equation}
which is also invariant under the same set of BRST transformations (\ref{brs}).
\section{$N=1$ Abelian super-Chern-Simons model in BV formulation}
In this section, we establish the theory in BV formulation.
For this purpose, we need antifields corresponding to fields  having opposite statistics.
In  terms of field and antifields, the generating functional  for the $N=1$ Abelian super-Chern-Simons theory  in Landau gauge  is defined  by  
\begin{eqnarray}
Z_L  = \int {\cal D}\Phi\ e^{ i\left(\Sigma_{inv}+ \Gamma^{\alpha\star} D_\alpha C +\widehat{C}^\star 
B \right) },\label{exl}
\end{eqnarray}
where $\Gamma^{\alpha\star}$ and $\widehat{C}^\star $ are antifields  corresponding to the 
$\Gamma^{\alpha}$ and $\widehat{C}$ fields with opposite statistics.
The above generating functional can further be written in compact form as
 \begin{equation}
Z_L = \int {\cal D}\Phi\  e^{ i    W_{\Psi^L  }[\Phi,\Phi^\star] },\label{lan}
\end{equation} 
where $ W_{\Psi^L}[\Phi,\Phi^\star] $ is an extended quantum action for the $N=1$ Abelian 
super-Chern-Simons theory in the Landau gauge
and   $\Phi^\star $ refers to the antifields generically corresponding to the collective field  $\Phi( \equiv \Phi_{\pm}, \bar\Phi_{\pm}, \Gamma_\alpha, C,  \widehat{C}, B)$.
The extended quantum action, $W_{\Psi  }[\Phi,\Phi^\star] $, satisfies a certain rich mathematical
relation, the so-called  quantum master equation \cite{wei},  which is given by
\begin{equation}
\Delta e^{iW_{\Psi }[\Phi,\Phi^\star] } =0,\ \
 \Delta\equiv \frac{\partial_r}{
\partial\Phi}\frac{\partial_r}{\partial\Phi^\star } (-1)^{\epsilon
+1}.
\label{mq}
\end{equation}
In other words, the extended quantum action $W_{\Psi }$ is the solution of the quantum master equation.
The antifields   $\Phi^\star $  for a general gauge theory can be evaluated from the expression of the gauge-fixed fermion.  For the $N=1$ super-Chern-Simons  theory in Landau  gauge the antifields are 
computed  with the help of the following gauge-fixed fermion $\Psi^L=\widehat{C}D^\alpha\Gamma_\alpha$:
 \begin{eqnarray}
\Phi_{1\pm}^{ \star }&=&\frac{\delta\Psi^L }{\delta \Phi_{\pm}}=0,
\nonumber\\
 \bar\Phi_{1\pm}^{ \star}&=&\frac{\delta\Psi^L }{\delta \bar\Phi_{\pm}}= 0,\nonumber\\ \Gamma_{1\alpha}^{\star}&=&\frac{\delta\Psi^L}{\delta \Gamma^\alpha}= -D^\alpha\widehat{C},\nonumber\\ \widehat{C}_1^{\star}&=&\frac{\delta\Psi^L}{\delta \widehat{C}}= D^\alpha\Gamma_\alpha,\nonumber\\ C^{\star}_1&=&\frac{\delta\Psi^L}{\delta C}=0.
\end{eqnarray}
With these values of antifields the extended quantum action in (\ref{exl}) coincides with
the total effective action (\ref{effl}).
However, the gauge-fixing fermion for the non-linear gauge choice  is given by
\begin{equation}
\Psi^{NL}=\widehat{C}(D^\alpha\Gamma_\alpha +\beta\Gamma_\alpha\Gamma^\alpha ).
\end{equation} 
The antifields with the help of the above gauge-fixing fermion for the
 non-linear gauge are calculated as:
 \begin{eqnarray}
\Phi_{2\pm}^{\star }&=&\frac{\delta\Psi^{NL} }{\delta \Phi_{\pm}}=0,
\nonumber\\
 \bar\Phi_{2\pm}^{ \star}&=&\frac{\delta\Psi^{NL} }{\delta \bar\Phi_{\pm}}= 0,\nonumber\\ \Gamma_{2\alpha}^{\star}&=&\frac{\delta\Psi^{NL}}{\delta \Gamma^\alpha}= -D^\alpha\widehat{C}+2\beta
 \widehat{C}\Gamma_\alpha,\nonumber\\ \widehat{C}_2^{\star}&=&\frac{\delta\Psi^{NL}}{\delta \widehat{C}}= D^\alpha\Gamma_\alpha
 +\beta\Gamma_\alpha\Gamma^\alpha,\nonumber\\ C_2^{\star}&=&\frac{\delta\Psi^{NL}}{\delta C}=0.
\end{eqnarray}
Similar to the linear gauge case, the  generating functional for the $N=1$ Abelian Super-Chern-Simons theory in non-linear gauge 
can   be written in compact form as
 \begin{equation}
Z_{NL} = \int {\cal D}\Phi\  e^{ i    W_{\Psi^{NL}  }[\Phi,\Phi^\star] },
\end{equation} 
where $ W_{\Psi^{NL}  }[\Phi,\Phi^\star] $ is an extended quantum action (a solution of the quantum master equation) in non-linear gauge.
\section{Solutions of the quantum master equation: field/antifield dependent symmetry}
In this section, we analyse the  field/antifield dependent 
BRST transformation which is characterized by the field/antifield dependent BRST  parameter.
To achieve the goal, we first define the usual BRST transformation for the generic field $\Phi_\alpha(x)$ written compactly as
 \begin{eqnarray}
\Phi_\alpha'(x)-\Phi_\alpha(x)=\delta_b  \Phi_\alpha(x)= s_b  \Phi_\alpha(x)\eta ={\cal R}_\alpha(x) \eta,
 \end{eqnarray}
where ${\cal R}_\alpha(x)(s_b  \Phi_\alpha(x))$ is the  Slavnov variation of the field $\Phi_\alpha(x)$ 
satisfying $\delta_b {\cal R}_\alpha(x)=0$.
Here the infinitesimal transformation  parameter $\eta$ is  a  Grassmann parameter.

Now, we propose the  field/ antifield dependent BRST transformation (as the discussed field-dependent BRST transformation in Ref. \cite{lav}) defined as
 \begin{eqnarray}
\delta_b  \Phi_\alpha(x)=\Phi_\alpha'(x)-\Phi_\alpha(x)={\cal R}_\alpha(x) \eta [\Phi,\Phi^\star],\label{qg}
 \end{eqnarray}
 where the Grassmann parameter   $\eta [\Phi,\Phi^\star]$  is the field/antifield dependent parameter of the transformation.
 The field/antifield dependent BRST transformation for the $N=1$ Abelian super-Chern-Simons theory is constructed by making the transformation parameter of (\ref{brs}) field/antifield dependent as
 \begin{eqnarray}
&&\delta_b\Phi_{\pm}=\pm iC\Phi_{\pm}\eta [\Phi,\Phi^\star],\ \  \ \delta_b\bar\Phi_{\pm}=\mp iC\bar\Phi_{\pm}\eta [\Phi,\Phi^\star],\nonumber\\
&&\delta_b\Gamma_\alpha =D_\alpha C\eta [\Phi,\Phi^\star],\ \ \ \ \ \ \delta_b\widehat{C}=B\eta [\Phi,\Phi^\star],\nonumber\\
&&\delta_b C=0,\ \ \ \ \ \ \  \delta_b B=0.\label{fi}
\end{eqnarray}
 Though the field/antifield dependent BRST transformation is not nilpotent in nature, it is
 the symmetry of the action. However it does not leave the generating functional invariant.
Under field/antifield dependent BRST transformation (\ref{fi}) the generating functional given in (\ref{lan}) transforms as 
\begin{eqnarray}
\delta_b Z_L &=& \int {\cal D}\Phi\ ( \mbox{sDet} J[\Phi, \Phi^\star] ) e^{ i    W_{\Psi^L  }[\Phi,\Phi^\star] },\nonumber\\
 &=&\int {\cal D}\Phi_i\  e^{i\left( W_{\Psi^L  }[\Phi,\Phi^\star]-i \mbox{Tr} \ln J[\Phi,\Phi^\star]\right)}.
 \label{zl} \end{eqnarray}
 Furthermore, we calculate the Jacobian matrix of the field/antifield dependent
BRST transformation   as  follows
 \begin{eqnarray}
J_\alpha^{\ \beta}[\Phi,\Phi^\star]= \frac{\delta \Phi'_\alpha}{\delta\Phi_\beta}&= &\delta_\alpha^{\ \beta}+
\frac{ \delta{\cal R}_\alpha(x)}{\delta\Phi_\beta} \eta[\Phi,\Phi^\star] +
{\cal R}_\alpha(x)\frac{ \delta\eta[\Phi,\Phi^\star]}{\delta\Phi_\beta},\nonumber\\
&= &\delta_\alpha^{\ \beta}+
{\cal R}_\alpha^{\ ,\beta}(x)  \eta[\Phi,\Phi^\star] +
{\cal R}_\alpha(x)\eta^{,\beta}[\Phi,\Phi^\star].\label{jac}
 \end{eqnarray}
Making use of the nilpotency property of the BRST transformation (i.e. $s_b^2=0$) and  (\ref{jac}),
we compute 
 \begin{equation}
 \mbox{sTr} \ln J[\Phi,\Phi^\star]=-\ln (1+s_b \eta[\Phi,\Phi^\star]),\label{J}
\end{equation}
where  $\eta[\Phi, \Phi^\star]$ exists up to linear orders only because of its anticommuting nature. 
Therefore, the Jacobian of  the arbitrary field/antifield dependent 
BRST transformation is given by
\begin{eqnarray}
\mbox{sDet} J[\Phi, \Phi^\star] = \frac{1}{1+s_b \eta[\Phi,\Phi^\star]}.
\end{eqnarray}
Now, with this identification of the Jacobian the expression (\ref{zl}) simplifies as
 \begin{eqnarray}
\delta_b Z_L  =\int {\cal D}\Phi\  e^{i\left( W_{\Psi^L  }[\Phi,\Phi^\star] +i\ln (1+s_b \eta[\Phi,\Phi^\star]) \right)},\label{dzl}
 \end{eqnarray}
which is nothing but the generating functional for the
super-Chern-Simons theory  having extended action $W_{\Psi^L  }[\Phi,\Phi^\star] +i\ln (1+s_b \eta[\Phi,\Phi^\star])$ where the extra piece is due to the Jacobian 
contribution.
We specifically choose 
the field/antifield  dependent transformation parameter
as follows
\begin{eqnarray}
\eta[\Phi,\Phi^\star]&=& \widehat{C}B^{-1}\left(e^{-is_b[\widehat{C} (\widehat{C}_2^\star-\widehat{C}_1^\star)]}  -1\right).
\end{eqnarray}
Now, we calculate the Jacobian contribution for this choice of the  field-dependent transformation parameter, which leads to
\begin{eqnarray}
-\ln(1+s_b\eta[\Phi,\Phi^\star]) &=&i[s_b \widehat{C} (\widehat{C}_2^\star-\widehat{C}_1^\star) +
\widehat{C} (s_b\widehat{C}_2^\star -s_b\widehat{C}_1^\star)].
\end{eqnarray}
Inserting the above value in (\ref{dzl}) we get,
\begin{eqnarray}
\delta_b Z_L  &=&\int {\cal D}\Phi\  e^{i\left( W_{\Psi^L  }[\Phi,\Phi^\star] +s_b \widehat{C} (\widehat{C}_2^\star-\widehat{C}_1^\star) +
\widehat{C} (s_b\widehat{C}_2^\star -s_b\widehat{C}_1^\star) \right)},\nonumber\\
  &=&\int {\cal D}\Phi\  e^{i\left( W_{\Psi^L  }[\Phi,\Phi^\star] +\beta 
  B\Gamma_\alpha\Gamma^\alpha +2\beta{\widehat C}\Gamma_\alpha D^\alpha  C \right)},\nonumber\\
  &=&\int {\cal D}\Phi\  e^{i  W_{\Psi^{NL}  }[\Phi,\Phi^\star]  },\nonumber\\
  &=&Z_{NL}.
 \end{eqnarray}
 Thus, we conclude that under field/antifield dependent BRST transformation with an appropriate choice of 
 field/antifield dependent parameter the generating functionals in linear and non-linear gauges
 are connected. In other words, we say that
 under field/antifield dependent BRST transformation the different solutions of quantum master
 equation can be connected. We established the results at the quantum level by using the BV formulation.
\section{  Conclusions}
In this paper we have considered the gauge invariant model of the $N=1$ Abelian Super-Chern-Simons 
theory and have analysed the theory at quantum level in different gauges, namely, in Landau and non-linear (quadratic)
gauges. The nilpotent BRST transformations are demonstrated for the effective actions
corresponding to both gauges. Furthermore we have analysed the theory at quantum level  in BV formulation which admits a mathematically rich
quantum master equation. The extended quantum actions are the solutions of 
such quantum master equation. Furthermore, we developed the field/antifield dependent 
BRST transformation characterized by the field/antifield dependent parameter.
For the field/antifield dependent BRST transformation we have calculated the Jacobian matrix explicitly.
Remarkably, we have found that under 
field/antifield dependent BRST transformation with an appropriate choice of the
transformation parameter the different solutions of the quantum master equation can be related.
We have shown the results by connecting the extended quantum actions in Landau and non-linear gauges.
Such results will help in clarifying the understanding of the dynamics of the theory in different gauges.

\end{document}